\documentclass[sigconf,nonacm]{acmart}

\begin{document}

\title{How Web Research Agents Interpret User Priorities}

\author{Robin Sharma}
\affiliation{%
  \institution{Independent Researcher}
  \city{Burlington}
  \state{VT}
  \country{USA}}
\email{sharmarolein@gmail.com}

\author{Parisa Suchdev}
\affiliation{%
  \institution{University of Vermont}
  \city{Burlington}
    \state{VT}
  \country{USA}}
\email{psuchdev@uvm.edu}

\renewcommand{\shortauthors}{Sharma and Suchdev}

\begin{abstract}
 Web research agents gather information online and write cited reports, including financial analysis. Users can state a priority to guide this research, but a priority may not say how to judge the options, leaving the agent to choose the criteria. We examine how two such agents, built on OpenAI and Anthropic models, interpret a stated priority. In 36 runs, the agents compared Microsoft and Alphabet as businesses. Prompts asked for special attention to growth or resilience, or gave no priority. We analyzed reports, searches, returned sources and cited sources. Of the 24 reports given a priority, 21 focused on it, but their criteria varied. Under the growth priority, OpenAI reports that ranked the companies accounted for risk or predictability and favored Microsoft, while the Anthropic ones judged growth by its pace and favored Alphabet. Under the resilience priority, reports used partly different criteria, and most named Microsoft as the more resilient company. Of the search and source measures tested, only the Anthropic agent's search wording differed detectably by condition. These findings suggest evaluations should examine the criteria behind a report's judgements, not only its focus or accuracy. Interfaces could let users see and adjust those criteria.
\end{abstract}

\begin{teaserfigure}
  \centering
  \includegraphics[width=\textwidth]{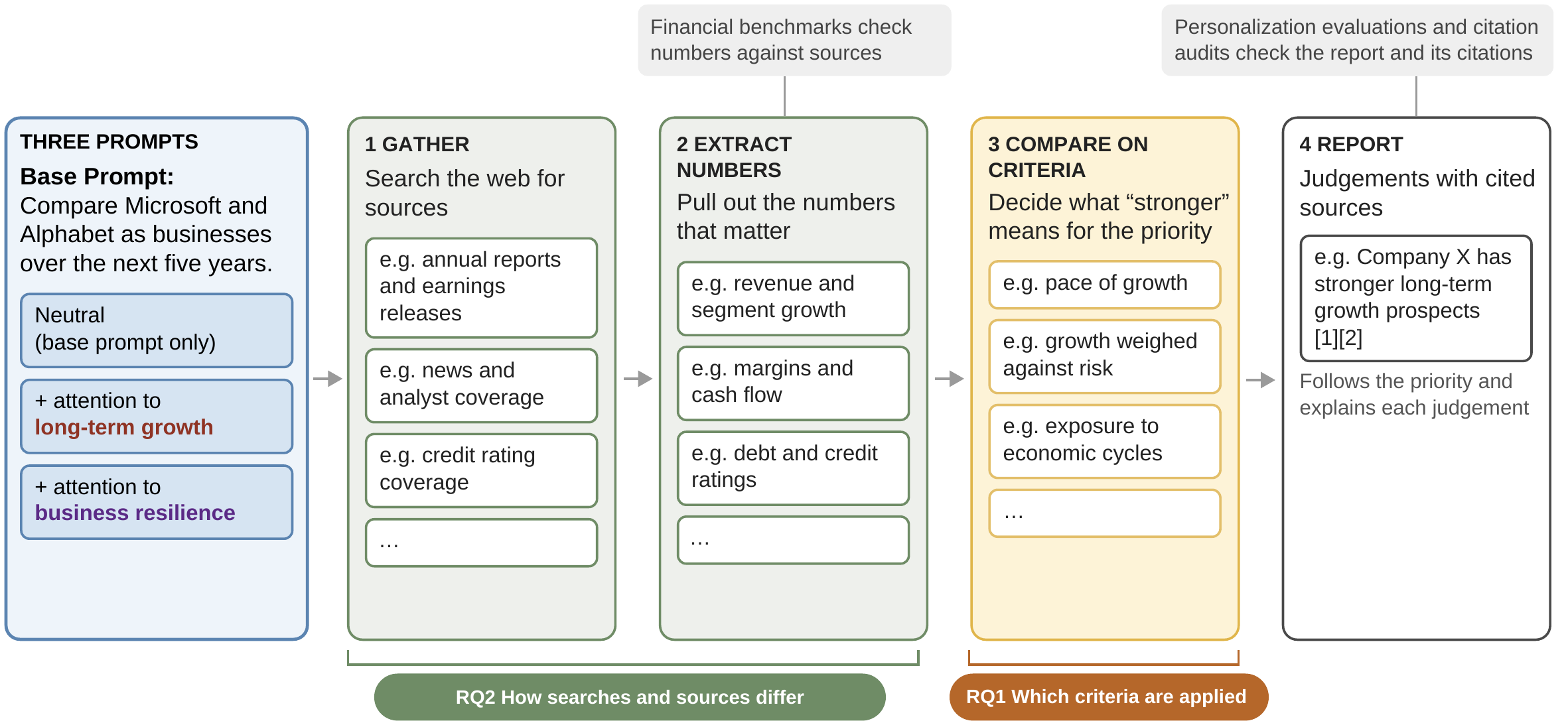}
  \caption{A conceptual illustration of a web research agent comparing two companies. The prompt names a priority but not how to judge it, so the agent decides which criteria to apply (RQ1) and what evidence to gather (RQ2). Existing evaluations check the numbers (step 2) and the final report (step 4), but not the criteria chosen in step 3.}
  \Description{A left-to-right flow diagram with five boxes. The first box is the prompt, which asks to compare Microsoft and Alphabet as businesses over the next five years, in one of three versions: with added attention to long-term growth, with added attention to business resilience, or with nothing added. Arrows lead to four steps. Step 1, gather, searches the web for sources such as annual reports and earnings releases, news and analyst coverage, and credit rating coverage. Step 2, extract numbers, pulls out figures such as revenue and segment growth, margins and cash flow, and debt and credit ratings. Step 3, compare on criteria, decides what stronger means for the priority, with examples such as pace of growth, growth weighed against risk and exposure to economic cycles. Step 4, report, gives judgements with cited sources, follows the priority and explains each judgement. Notes above step 2 and step 4 say that financial benchmarks check numbers against sources and that personalization evaluations and citation audits check the report and its citations. Step 3 has no note. A bracket under steps 1 and 2 is labeled RQ2, how the priority shapes the evidence, and a bracket under step 3 is labeled RQ1, which criteria are applied.}
  \label{fig:concept}
\end{teaserfigure}
\maketitle
\section{Introduction}
Large language models (LLMs) can struggle with questions that require up-to-date information \cite{vu2023freshllms}. Web search gives them access to external sources \cite{nakano2021webgpt}. Agent-based systems go further and let the model decide what to search for, which pages to inspect and when it has enough information to answer \cite{anthropic2024building}. We call such systems web research agents when they search the web and produce cited reports~\cite{bosse2025deep}.

Two users can give a web research agent the same task and still need different reports, because they have different goals \cite{yoon2026personalized, li2026personalized}. A user can tell the agent what matters by stating a priority, asking it to give special attention to one aspect of the task. A stated priority can tell the agent what to pay attention to while leaving open how to judge the options. Decision analysis treats these as separate choices, choosing an objective and deciding how to measure performance on it \cite{keeney2005selecting}. We call the grounds a report gives for comparing options on a priority its assessment criteria. When the user does not specify these grounds, the agent has to choose them, and its choice could change which option the report recommends.

We study this problem with a financial task, a comparison of two companies. Financial decisions are high-stakes because they involve people's money, and LLM-based advisors are becoming increasingly popular as a low-cost source of personalized financial advice \cite{takayanagi2025generative}, even though LLMs can favor particular stocks and types of companies \cite{lee2025your, zhi2025exposing}. Financial priorities also have several reasonable criteria. A user interested in how much the businesses could expand might ask for special attention to long-term growth potential. A report could then compare the companies by their expected growth rates or weigh those rates against downside risk. A user interested in how well the businesses could withstand difficult periods might instead prioritize long-term business resilience. A report could then compare their exposure to economic and advertising downturns or their credit ratings, which grade how likely each company is to repay its debt.

One way to settle these criteria is to learn more about the user. Personalized research systems ask clarifying questions, refine the request using information about the user or build a profile from the user's documents \cite{luo2026intentrl, yoon2026personalized, li2026personalized}. Benchmarks, by contrast, set their own criteria and score reports against them \cite{du2026deepresearch, haque2026deepfinresearch, zhu2025findeepresearch}, which shows how well a report meets those criteria but not which criteria the agent applied. Less is known about what an agent does when the prompt specifies a task and a priority but doesn't say how to compare the options. In that case, its report can give the priority special attention, and so appear to follow the request, while supplying grounds for comparison that the prompt left open.

A priority may also change what the agent investigates, not only what its report says. An agent might search for different information and find different sources. Even if it finds the same sources, it might look for different details within them or cite different ones in the report. In a related setting, DRaligned found that differences between user conditions appeared more clearly in final recommendations than in the agents' information requests (their search queries and requests to read pages) \cite{liu2026happens}. Knowing whether a priority changes the research, and in which part, tells evaluators which parts of a run to examine and tells designers where to show users how their priority was used.

We studied two web research agents, one built on an OpenAI model and one on an Anthropic model. We accessed each model through its provider's API with web search enabled. The base prompt asked both agents to compare Microsoft and Alphabet as businesses over the next five years and to support their analysis with cited sources. The growth and resilience conditions each added one sentence asking for special attention to that priority. The neutral condition used the base prompt alone, allowing us to see which criteria also appeared without a stated priority. Each agent ran each condition six times, for 36 runs with no further user input. We recorded whether each report focused on growth, focused on resilience or balanced the two. We then examined whether it named either company as stronger and what reasons it gave for its comparisons. We also compared search queries, returned sources and cited sources across conditions. We ask two research questions.
\begin{itemize}
    \item RQ1 (Interpretation). How do web research agents interpret a stated priority, and what criteria do their reports give for comparing the companies on it?
    \item RQ2 (Research process). How do the agents' searches and sources differ across priority conditions?
\end{itemize}

No agent asked what the priority meant, and every run ended with a report. Although 21 of the 24 reports given a priority adopted the requested focus, the criteria behind their comparisons varied, and some also appeared without a priority (RQ1). Among growth reports that named a stronger company, the OpenAI reports weighed growth against risk or predictability and favored Microsoft, while the Anthropic reports judged growth by how fast each company was growing and favored Alphabet. Under the resilience priority, all nine reports that named a more resilient company chose Microsoft, though the grounds they gave partly differed. For RQ2, the Anthropic agent searched with different words under different priorities, and in four of its resilience runs credit ratings appeared in its searches, its citations and its comparison of the companies. The OpenAI agent's searches and sources showed no detectable difference by condition, although individual runs looked up different figures within the pages they opened.
We make three contributions.
\begin{itemize}
    \item A study design that compares runs with and without a stated priority across their reports, searches and sources.
    
    \item Evidence that reports usually adopted the requested focus. Under the growth priority, the reports that ranked the companies favored different companies on different grounds (RQ1). Among the search and source measures tested, only the Anthropic agent's search wording changed by condition (RQ2).
    
    \item Implications for evaluating web research agents and for interfaces that let users see and change the criteria an agent uses.
\end{itemize}

\section{Related Work}
\subsection{ Instruction Following, Clarification and Personalization}
Instruction-following research examines whether language models satisfy stated requirements. IFEval focuses on instructions that can be checked automatically, since compliance is harder to verify for requests such as writing in a funny tone \cite{zhou2023instruction}. Work on intent hallucination examines cases where a model omits or misinterprets part of a request \cite{hao2025beyond}. When a request leaves information unstated, search agents can ask for clarification. InteractComp finds that the agents it tested often answered before gathering enough information \cite{deng2026interactcomp}, and DiscoBench tests whether agents can find the meaning a user intended when a request leaves the ranking criteria unclear \cite{tao2026discobench}. Personalization studies assess how well responses reflect information needs users describe~\cite{kim2026bespoke} or how closely reports follow specified research actions \cite{balepur2026language}.
\begin{figure*}[t]
  \centering
  \includegraphics[width=\textwidth]{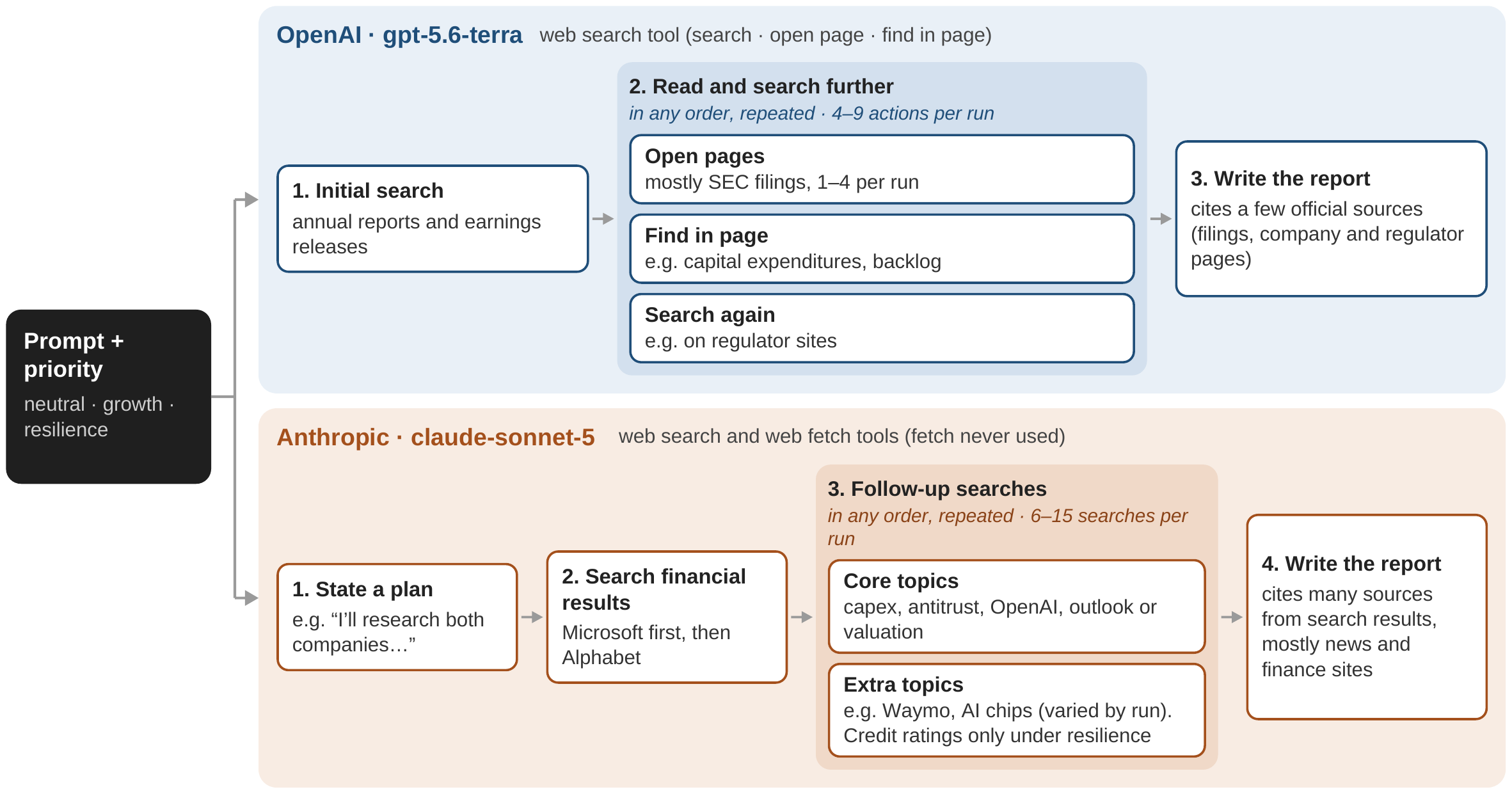}
  \caption{Typical research steps of each agent across its 18 runs. Numbered steps occurred in this order in every run. Steps stacked inside a shaded group occurred in any order and could repeat within a run. The OpenAI agent opened official documents and searched within them, while the Anthropic agent wrote its report from search results without opening any pages.}
  \Description{A flow diagram with a prompt box on the left that branches into two horizontal lanes, one per agent. The prompt box reads prompt plus priority, with the conditions neutral, growth and resilience. The top lane, OpenAI gpt-5.6-terra, uses one web search tool that can search, open a page and find in a page. Its steps are 1, initial search for annual reports and earnings releases. 2, read and search further, a group of three stacked actions done in any order and repeated, 4 to 9 actions per run. The actions are open pages, mostly SEC filings, 1 to 4 per run, find in page, for example capital expenditures and backlog, and search again, for example on regulator sites. 3, write the report, citing a few official sources such as filings, company pages and regulator pages. The bottom lane, Anthropic claude-sonnet-5, has web search and web fetch tools, and fetch was never used. Its steps are 1, state a plan, for example I will research both companies. 2, search financial results, Microsoft first, then Alphabet. 3, follow-up searches, a group of two stacked items done in any order and repeated, 6 to 15 searches per run. The items are core topics, namely capex, antitrust, OpenAI, and outlook or valuation, and extra topics that varied by run, for example Waymo and AI chips, with credit ratings only under resilience. 4, write the report, citing many sources from search results, mostly news and finance sites.}
  \label{fig:workflow}
\end{figure*}
These evaluations check responses against stated instructions, the meaning a user intended or the needs and research actions users specified. A stated priority can be followed in several reasonable ways, as the examples in the introduction show. Choosing among those ways differs from omitting or misinterpreting a stated requirement. This leaves open which criteria reports use to justify their comparisons when the user gives a stated priority and nothing more.

\subsection{Evaluation of Research Outputs and Processes
}
Benchmarks for web research agents assess outputs ranging from short answers to full reports. FinSearchComp evaluates answers to expert-written financial questions~\cite{hu2025finsearchcomp}. DeepResearch Bench scores reports against task-specific criteria and separately measures citation accuracy \cite{du2026deepresearch}. Financial report benchmarks use detailed rubrics~\cite{zhu2025findeepresearch} or compare generated reports with professional analyst reports, assessing qualities such as how explicitly the reports state and justify their assumptions \cite{haque2026deepfinresearch}. Several of these benchmarks test both commercial research products and general-purpose models equipped with web search \cite{du2026deepresearch, haque2026deepfinresearch, zhu2025findeepresearch}. Our agents belong to the second group. These benchmarks score outputs against evaluation standards, whereas we identify the criteria reports use to justify their comparisons.

Other work examines how systems that search the web find and use evidence. Citation audits test whether cited sources support the claims they accompany \cite{liu2023evaluating, venkit2026deeptrace}. DeepHalluBench reports that the agents it tested favored evidence encountered early or repeatedly \cite{zhan2026why}. Search Shapes Conclusions examines whether the evidence in the documents an agent reads reflects the evidence in a predefined pool of candidate documents \cite{xiao2026search}. DRaligned examines how a change in user context is reflected in agents' information requests, working drafts and final recommendations \cite{liu2026happens}. We compare the queries agents issued, the sources those queries returned and the sources their reports cited across priority conditions.

\subsection{ Explainability and User Control}
\citet{shah2022situating} argue that presenting a single answer shifts decisions about which information matters and how to interpret it from the user to the search system. In a study of answer engines, participants clicked few citations, and responses to questions that implied a preferred position often reinforced that position \cite{venkit2025search}. When a web research agent compares alternatives, readers may need to see the criteria for the comparison and the evidence supporting it. Research on explainable AI suggests designing explanations around users' questions \cite{liao2020questioning}, including why one outcome occurred instead of another \cite{miller2019explanation}. In \citeauthor{amershi2019guidelines}'s study \cite{amershi2019guidelines}, a participant complained that a navigation app called its route the ``best route'' without saying what made it the best. A reader of a research report can ask the same question about the option the report favors.

The question matters in financial research, where reports can inform investment decisions and systems such as FinSight now generate them \cite{jin2025finsight}. In a lab study of LLM financial advisors, participants' perceptions of the advisors did not consistently track how closely their asset rankings matched expert judgements \cite{takayanagi2025generative}. FINRA identifies the difficulty of tracing outcomes from multi-step agent reasoning as a challenge for auditing~\cite{finra2025oversight}. A report can state the basis for its judgement, and citations let readers inspect the evidence it presents.

Research interfaces give users ways to inspect and steer an agent's work. InterDeepResearch lets users trace claims to supporting evidence \cite{pan2026interdeepresearch}. MyScholarQA lets users review proposed research actions and highlights passages associated with the actions they approve \cite{balepur2026language}. EvalLM lets users define criteria for evaluating model outputs \cite{kim2024evallm}. People can struggle to express their requirements in prompts \cite{zamfirescu2023johnny}, and evaluating outputs can lead them to revise their criteria \cite{shankar2024validates}. These tools show evidence, proposed actions or criteria that users define. We examine the criteria a report applies when the user has not defined them, which an interface could show and let users change.

\section{Study Design and Analysis}

\subsection{ Task, Conditions and Agents}
We chose Microsoft and Alphabet because both are documented in detail in annual filings, earnings releases and news coverage. Every prompt began with the same base text, ``Research and compare Microsoft and Alphabet as businesses, considering their prospects over the next five years. Support your analysis with cited sources.'' The growth condition added ``In your research, give special attention to long-term growth potential.'' The resilience condition added ``In your research, give special attention to long-term business resilience.'' The neutral condition used the base text alone. The prompts left open how to judge growth or resilience, which sources to consult and which company to favor. The added sentence referred to the research, so it asked for attention to the priority in the searches as well as in the report.

Both agents ran in their provider's built-in tool loop, and we supplied no system prompt. The OpenAI agent used gpt-5.6-terra with medium reasoning effort and OpenAI's web search tool, which could run queries, open pages and search within them. The Anthropic agent used claude-sonnet-5 with high reasoning effort and had access to web search and web fetch, but it made no web fetch calls (Appendix A). Figure 2 summarizes how each agent conducted its research. Because the agents differ in their models, tools and settings, we compare conditions within each agent and report differences between agents without attributing them to the model.

To compare differences between conditions with the variation between repeated runs of the same prompt, we ran each condition six times per agent (36 runs). We submitted one batch per provider after US markets had closed. All results were saved within about six minutes.

\subsection{ Recorded Data and Labeling
}
For each run, we saved the agent's search queries, the sources each search returned, the sources its report cited and the report itself. A query is one search string, such as ``Microsoft Alphabet credit rating AAA Moody's S\&P 2025'', and each distinct URL counted as a source. Returned sources were those listed in search results, and cited sources were those linked in the report. Anthropic citation records also included excerpts from the cited sources, but we identified cited sources by URL for both agents so that the analysis used the same unit throughout. For the OpenAI agent, we also saved its page openings and its searches within those pages. Appendix A gives the remaining recording details.

\begin{table*}
  \caption{Labels used in the analysis.}
  \label{tab:labels}
  \begin{tabular*}{\textwidth}{@{\extracolsep{\fill}}llp{0.28\textwidth}p{0.32\textwidth}@{}}
    \toprule
    Label & Applied to & Values & Records \\
    \midrule
    Report orientation & Reports & \raggedright Growth, resilience, balanced or unclear & \raggedright The report's focus, the priority that organized it (RQ1) \tabularnewline
    Overall preference & Reports & \raggedright Microsoft, Alphabet, no overall choice stated or unclear & \raggedright Which company the report favored overall (RQ1) \tabularnewline
    Growth comparison & Reports & \raggedright Microsoft stronger, Alphabet stronger, not stated or unclear & \raggedright Which company the report judged stronger on growth (RQ1) \tabularnewline
    Resilience comparison & Reports & \raggedright Microsoft stronger, Alphabet stronger, not stated or unclear & \raggedright Which company the report judged more resilient (RQ1) \tabularnewline
    Microsoft safer / predictable & Reports & \raggedright Yes, no or unclear & \raggedright Whether the report called Microsoft safer or more predictable (RQ1) \tabularnewline
    Alphabet higher upside & Reports & \raggedright Yes, no or unclear & \raggedright Whether the report said Alphabet had more upside (RQ1) \tabularnewline
    Creditworthiness & Reports & \raggedright Yes, no or unclear & \raggedright Use of creditworthiness to support an assessment (RQ1, RQ2) \tabularnewline
    Growth (0/1) & Queries & \raggedright 0 or 1 & \raggedright Whether the query asked about growth (RQ2) \tabularnewline
    Resilience (0/1) & Queries & \raggedright 0 or 1 & \raggedright Whether the query asked about resilience (RQ2) \tabularnewline
    \bottomrule
  \end{tabular*}
\end{table*}
The first author (Coder 1) and the second author (Coder 2) labeled the reports and search queries with the labels in Table 1, without seeing the agent, run or condition that produced them. Coder 1 labeled all 36 reports. Report orientation recorded which priority organized a report's analysis and conclusion, which we call the report's focus. A report adopted the requested focus when its focus matched the priority in its prompt. We labeled a report balanced when it developed both priorities and drew on both in its conclusion, without favoring either. The next five labels counted only what a report stated explicitly. Overall preference, growth comparison and resilience comparison recorded which company, if any, a report named as the better business, as stronger on growth or as more resilient.

Reports also compared the companies without ranking them. One OpenAI report, for example, said that Microsoft had ``the more predictable five-year earnings architecture'' while Alphabet had ``the larger strategic upside'' (resilience run 01). Because the comparison labels record only rankings, they miss judgements like this one. Two further labels therefore recorded whether a report called Microsoft safer or more predictable and whether it said Alphabet had more upside, the two descriptions in this example. We applied them to every report, including reports that also named a stronger company, so we could see how often each description appeared in each condition. We also searched all 36 reports for any report that called Alphabet safer or said Microsoft had more upside. Finally, the creditworthiness label recorded whether a report used a credit rating, credit standing, solvency or default risk to support an assessment of either business. We labeled this criterion so that we could trace it through each run's research (Section 3.3). A rating that a report mentioned without using it to judge the companies did not count, because the label recorded the criterion, not the information.

Coder 1 also labeled all 450 search queries, shown without the agent, run or condition that produced them. Each query received a growth label and a resilience label separately, so it could receive both or neither, and neither label required the words growth or resilience. A query counted as growth when it asked about growth or expansion, such as revenue growth, adoption, monetization, new markets or market-share gains, and past growth also counted. Static figures, valuation or general background did not count. A query counted as resilience when it asked about sustaining performance, withstanding adverse conditions or downside exposure, such as business risks, regulatory exposure, disruption, dependence on partners, cyclicality, credit ratings or how durable revenue is. A topic that could affect durability did not count unless the query asked about it in those terms.

To check the reliability of these labels, Coder 2 independently labeled a random sample. The sample contained 8 of the 36 reports, four per agent and at least one from each of the six agent and condition combinations, and 90 of the 450 queries (20\%), sampled at random within each run. Coder 2 applied the six report labels other than creditworthiness, which only Coder 1 labeled, and both query labels.

Coder 1 and Coder 2 agreed on 45 of the 48 report labels, including 7 of the 8 report orientation labels, and resolved the disagreements through discussion. For the sampled queries, Cohen's $\kappa$~\cite{cohen1960coefficient} was 1.00 for growth and .84 for resilience. The discussion clarified the rule for resilience queries, which Coder 1 reapplied to all 450 queries, and the analysis used Coder 1's revised labels. Appendix C gives the earlier pilot and the full coding rules.

\subsection{Analysis}\label{sec:analysis}
For RQ1, we grouped the reports by agent and condition and counted how many focused on growth, on resilience or on both equally, and which company they favored. We also counted how many called Microsoft safer and how many said Alphabet had more upside. We then identified the assessment criteria reports gave for comparing the companies on growth or resilience, such as how fast each company was growing or how exposed it was to economic downturns. The two coders identified these criteria together. They first read the 17 reports that explicitly named one company as stronger on growth or more resilient. One report ranked the companies on both, bringing the total to 18 rankings. For each ranking, they read the passages that gave its reasons and recorded the criteria behind them. Reports could also use a criterion without ranking the companies, so the coders searched all 36 reports for words that signal a criterion, such as ``risk-adjusted'', ``predictable'', ``cyclical'' and ``credit rating'', and read every match. We tracked two resilience criteria across all reports because the agents' resilience rankings differed on them. The first was exposure to economic and advertising cycles, a standard part of how credit rating agencies assess business risk \cite{spglobal2019corporate}, counted only when a report compared the two companies on it. The second was creditworthiness, taken from Coder 1's label (Section 3.2). All four reports with this label used credit ratings, so we refer to it as credit ratings. Appendix D lists the rankings and criteria we recorded by run.

For RQ2, we asked whether the agents' searches and sources differed across priority conditions. For each agent, we compared every pair of runs on three things, the words in their search queries, the URLs their searches returned and the URLs their reports cited. Before comparing wording, we removed common words and selected terms related to the priorities. We measured how alike two runs were with the Jaccard index \cite{jaccard1901etude}, as HypoSearch did for search queries \cite{zhou2026explore}. It is the number of items the two runs share divided by the number of distinct items across both. We then computed the gap between the average similarity of runs given the same prompt and that of runs given different prompts. We tested this gap with a permutation test, shuffling the condition labels across runs within each agent and counting how often a gap at least as large as the observed one arose by chance. We made this comparison across all three conditions and for each pair of conditions. We compared report wording in the same way, which shows whether reports differed by condition without relying on the orientation labels.

Overlap can hide differences that involve only a few items, so we also used two other measures. Across all three conditions, we tested the number of URLs cited in at least two runs of one condition and in no run of another. For each pair of conditions, we compared the average share of each run's queries labeled growth or resilience. Both tests used the same label shuffling. Two further analyses were descriptive. We described the OpenAI agent's 22 searches within opened pages across its 18 runs. For each report that used credit ratings, we went back through its run to find the queries that asked about them, the cited sources that contained them and the passages that relied on them.

We report all 46 tests and adjust them together with Holm's method \cite{holm1979simple}, so that the chance of even one false positive across them is at most 5\%. We call a result significant only after this adjustment, and a result that is not significant means no difference was detected, not that the runs were the same. Appendix B gives the procedures and every result.
\begin{table*}
  \caption{Report focus and overall pick by condition. Each agent produced six reports per condition.}
  \label{tab:focus}
  \begin{tabular}{lllll}
    \toprule
    Condition & OpenAI focus & OpenAI overall pick & Anthropic focus & Anthropic overall pick \\
    \midrule
    Neutral & 6 balanced & Microsoft, 2 of 6 & 6 balanced & No overall pick \\
    Growth & 5 growth, 1 balanced & Microsoft, 3 of 6 & 4 growth, 2 balanced & No overall pick \\
    Resilience & 6 resilience & Microsoft, 1 of 6 & 6 resilience & No overall pick \\
    \bottomrule
  \end{tabular}
\end{table*}
\section{Findings}
We first examine whether reports adopted the requested focus (4.1), then the criteria they applied (4.2) and whether the agents' searches and sources differed across conditions (4.3).

\subsection{RQ1: Reports adopted the requested focus}
21/24 reports given a growth or resilience priority focused on it (Table 2). The other three were growth reports that showed no preference between growth and resilience. The 12 reports without a priority also showed no preference between the two. Comparing the reports' wording, which does not depend on our labels, supported this. For both agents, reports given the same priority used more similar wording than reports given different priorities (OpenAI adjusted $p = .035$, Anthropic adjusted $p = .018$). Only six reports named an overall pick, all from the OpenAI agent and all choosing Microsoft.

\subsection{RQ1: Reports used different criteria for the same priority}
Six reports under the growth priority named a company with stronger growth prospects (Figure 3). The four OpenAI reports accounted for risk or predictability when judging growth and ranked Microsoft higher. The two Anthropic reports judged growth by its pace and ranked Alphabet higher. Reports from each agent also acknowledged the other company's strength. An OpenAI report noted Alphabet's ``stronger reported growth acceleration'' (growth run 05), and an Anthropic report noted that Microsoft ``offers more downside protection'' (growth run 06).

Under the resilience priority, six OpenAI reports and three Anthropic reports named a more resilient company, and all nine chose Microsoft (Figure 4). Their reasons differed by agent. The OpenAI reports pointed to Microsoft's lower exposure to economic and advertising cycles, while the Anthropic reports pointed to its credit ratings. A fourth Anthropic report also used credit ratings but did not name a more resilient company.

Some criteria also appeared without the matching priority. One neutral and one resilience report from Anthropic ranked Alphabet higher on growth based on its pace. In each condition, most OpenAI reports described Microsoft as safer or more predictable, and every OpenAI report said Alphabet had more upside. No report called Alphabet safer or said Microsoft had more upside.
\begin{figure*}[t]
  \centering
  \includegraphics[width=\textwidth]{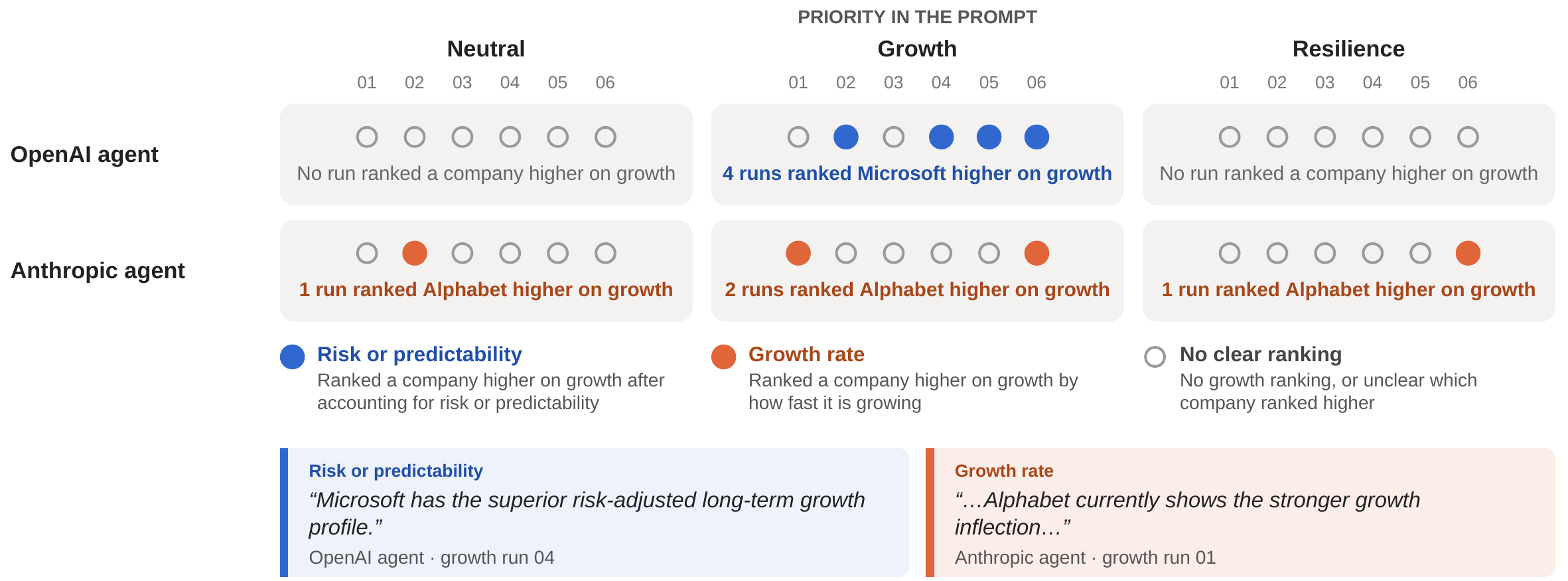}
  \caption{How each agent judged growth, by priority condition. Each mark is one report, and the numbers identify runs. A filled mark means the report ranked one company higher on growth, and its color shows how the report judged growth. The OpenAI agent ranked Microsoft higher on growth after accounting for risk or predictability, while the Anthropic agent ranked Alphabet higher based on how fast it was growing. The figure shows growth rankings only. Overall picks are reported in Table~\ref{tab:focus}.}
  \Description{A grid with two rows, one per agent, and three columns for the neutral, growth and resilience prompts. Each cell holds six marks for runs 01 to 06. In the OpenAI agent row, the neutral and resilience cells have no filled marks, and the growth cell has filled blue circles for runs 02, 04, 05 and 06, labeled 4 runs ranked Microsoft higher on growth. In the Anthropic agent row, filled orange squares appear for neutral run 02, growth runs 01 and 06, and resilience run 06, labeled as runs that ranked Alphabet higher on growth. A legend explains that a blue circle means growth was judged after accounting for risk or predictability, an orange square means growth was judged by how fast the company is growing, and an open ring means no clear growth ranking. Two quotes appear below. The OpenAI agent, growth run 04, wrote that Microsoft has the superior risk-adjusted long-term growth profile. The Anthropic agent, growth run 01, wrote that Alphabet currently shows the stronger growth inflection.}
  \label{fig:growth}
\end{figure*}

A user who asked for attention to growth could therefore receive Microsoft because its growth was steadier, or Alphabet because it was faster, depending on the agent. Almost every report that ranked the companies on the requested priority also adopted its focus (14 of 15). A report's focus therefore did not reveal which criteria it used.
\begin{figure*}[t]
  \centering
  \includegraphics[width=\textwidth]{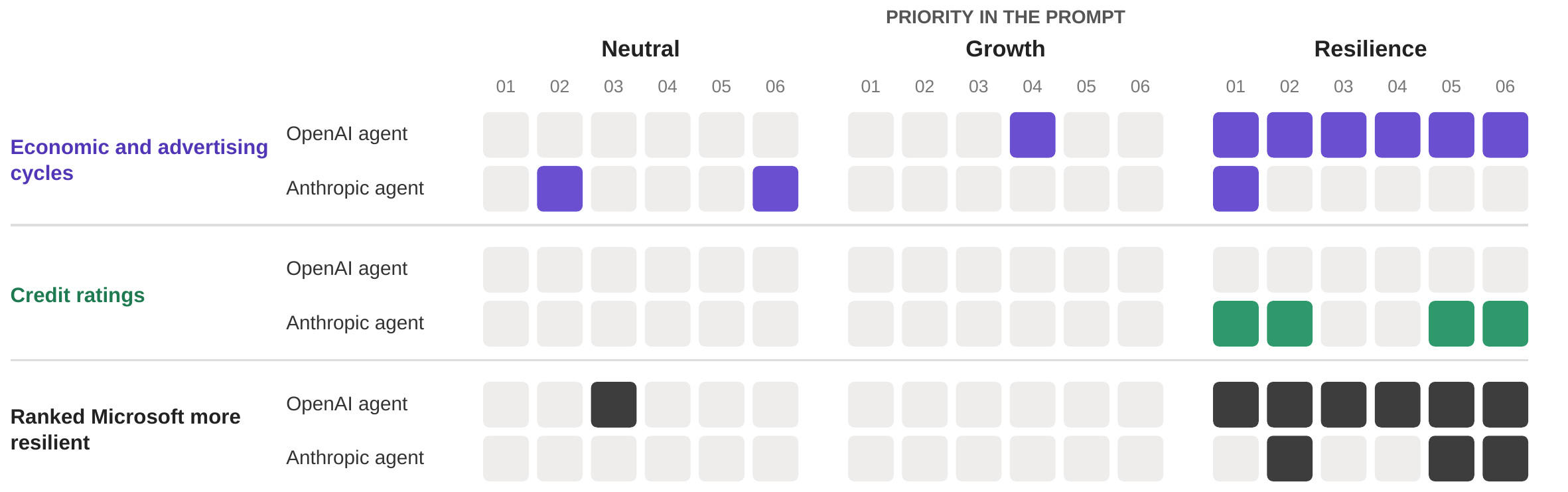}
  \caption{Criteria the reports used to compare resilience. Each square is one report. In the first two groups, a filled square means the report used that criterion, whether or not it named a more resilient company. Every OpenAI resilience report compared exposure to economic and advertising cycles, which only one of its other reports did. Four Anthropic resilience reports used credit ratings, which no other report mentioned. The bottom group shows the ten reports that ranked Microsoft more resilient, nine of them under the resilience priority. No report ranked Alphabet more resilient.}
  \Description{A grid of squares with three column groups for the priority in the prompt (neutral, growth and resilience). Each column group has six columns for runs 01 to 06. There are three row groups, each with one row for the OpenAI agent and one for the Anthropic agent. Economic and advertising cycles, in purple: the OpenAI agent is filled in all six resilience runs and in growth run 04, and the Anthropic agent is filled in neutral runs 02 and 06 and resilience run 01. Credit ratings, in green: the OpenAI agent has no filled squares, and the Anthropic agent is filled in resilience runs 01, 02, 05 and 06. Ranked Microsoft more resilient, in dark grey: the OpenAI agent is filled in neutral run 03 and all six resilience runs, and the Anthropic agent is filled in resilience runs 02, 05 and 06. All other squares are light grey.}
  \label{fig:resilience}
\end{figure*}

\subsection{RQ2: Searches and sources across priority conditions}
For the Anthropic agent, runs given the same priority shared more of their search words than runs given different priorities (41\% vs. 33\% average overlap, Table 3). The overall test across the three conditions remained significant after the Holm adjustment (adjusted $p = .009$), but no single pair of conditions differed after adjustment (adjusted $p = .09$ for each pair). The sources its searches returned and its reports cited did not differ detectably by condition (smallest adjusted $p = .24$). Its cited sources varied widely from run to run in every condition, with an average overlap of 5\% between pairs of reports.
\begin{table}
  \caption{Searches and cited sources by agent. The first four rows describe each agent's sources regardless of condition, and the last two compare conditions. Overlap is the average share of items two runs have in common (Section~\ref{sec:analysis}). The $p$ value is Holm-adjusted.}
  \label{tab:sources}
  \begin{tabular*}{\linewidth}{@{\extracolsep{\fill}}p{0.47\linewidth}ll@{}}
    \toprule
     & OpenAI & Anthropic \\
    \midrule
    \raggedright Sources cited per report (average) & 7 & 35 \\
    \raggedright Different sources cited across 18 runs & 21 & 391 \\
    \raggedright Citations to sources cited under all three conditions & 94\% & 24\% \\
    \raggedright Source overlap between two reports & 44\% & 5\% \\
    \raggedright Search-word overlap, same vs.\ different prompt & 43\% vs.\ 43\% & 41\% vs.\ 33\% \\
    \raggedright Search-word comparison across all three conditions, adjusted $p$ & 1 & .009 \\
    \bottomrule
  \end{tabular*}
\end{table}
For the OpenAI agent, runs given the same priority and runs given different priorities shared about the same proportion of search words (43\% vs. 43\%). Neither its search wording nor its cited sources differed detectably by condition (unadjusted $p = .52$ and $.49$, so not significant even before adjustment). Its reports drew on the same few sources in every condition, with 94\% of its citations going to nine documents, mostly the companies' own filings, that were cited under all three conditions. For both agents, the remaining tests, of the share of growth and resilience queries and of URLs cited repeatedly in only one condition, showed no detectable difference after adjustment (Appendix B).

The OpenAI agent's tool records show what it looked for within pages. It searched within opened pages 22 times across 16 of its 18 runs. Capital expenditures was the most common search term in every condition. A few runs in each condition searched for terms no other run used, such as additions to property and equipment in a growth run and backlog in a resilience run.

In each of the Anthropic agent's four reports that used credit ratings, the same criterion appeared in its searches, its cited sources and the report's judgement. In resilience run 05, for example, it searched for ``Microsoft Alphabet credit rating AAA Moody's S\&P 2025'', cited a page reporting Microsoft's AAA ratings from S\&P, Moody's and Fitch, and wrote that Microsoft ``holds the strongest possible credit profile in corporate America''. This shows that the criterion appeared at each step, not that the search caused the report's conclusion.

\section{Implications for Evaluation and Design}
\subsection{ Evaluate the criteria behind a report's comparison}
Checking what a report focuses on is not enough to evaluate how it handles a priority, because reports with the same focus compared the companies on different criteria. Evaluations should therefore record the criteria each report gives for its comparison and check how those criteria vary across repeated runs of the same prompt and across priorities. Larger studies could have a language model record these criteria, after first checking that its labels agree with human coders on a sample.

Once the criteria are recorded, the evidence behind each one needs checking too, and that evidence often rests on a specific figure or passage within a source. The OpenAI agent's tool records show which terms it looked for inside documents, and 20 of its 22 searches within pages were in the nine documents it cited in every condition. Its citation records, however, give only each document's URL, so they do not show which passage of a document supports a given sentence in the report. Evaluators could therefore record the passage in the report that gives each criterion and, where possible, the figure or passage in the cited source that supports it. Our credit-rating trace shows that searches, citations and report text can together link a criterion to its evidence.

Even a well-supported criterion may not be the one the user intended. Credit ratings concern debt repayment, and the reports that used them cited sources that stated the ratings. A user asking about resilience might care more about how each company's business would hold up in a recession. Report text alone cannot show which grounds the user would prefer.

\subsection{ Make the criteria for a priority visible and reviewable}
Before research begins, an agent could show users the criteria it proposes to apply. Users could then choose or revise them, an approval step similar to MyScholarQA's \cite{balepur2026language}. In the report, each judgement could name the criterion it used and indicate whether the user selected it or the agent proposed it. The report could link to the cited evidence that supports the comparison and, where the connection can be verified, show the queries that addressed the criterion. Users could also revisit these criteria after reading the report \cite{shankar2024validates}.

Users could choose to save criteria for similar future tasks. MyScholarQA's pilot survey found that users were willing to put in some effort up front and disliked restating their needs for each query~\cite{balepur2026language}. A research agent could build on approaches that reuse user context \cite{balepur2026language, li2026personalized} by making saved criteria visible and editable when they are suggested for a new task.

Reviewing criteria takes time, especially when users are unfamiliar with the choices, and specifying interests or requirements already takes effort \cite{chang2019searchlens, zamfirescu2023johnny}. A user study could test when reviewing criteria helps people get a comparison they find useful and how much effort they are willing to spend on that review.

\section{Limitations}
This is a small study. We ran one comparison task with two agents on a single day, and we stated each priority in a single sentence, such as ``give special attention to long-term growth potential.'' A different task, a differently worded priority or a newer model version could lead the agents to use different criteria. We also called the models through their APIs, and commercial Deep Research products may add their own instructions and tools and behave differently. Both companies are large technology firms, so comparisons involving smaller companies or other sectors are needed to see how far our findings apply.

Even within this setting, the tests could miss things. Six runs per agent and condition may have missed smaller differences, including any in the OpenAI agent's searches. The word comparisons count shared words rather than meaning, so two reports could say the same thing in different words, or different things in similar words. Although we adjusted for the number of tests, the adjustment does not cover choices we made after looking at the data, such as which criteria to track across all reports.

The labels have limits too. Because they counted only explicit statements, a report could rely on a criterion without stating it and go uncounted. The two coders identified criteria together, so that step had no independent check, and the credit-rating finding rests on labels from Coder 1 alone. 

Last, the records show what the agents did, not why. They cannot tell us why an agent chose one criterion over another, and the criteria a report states may not include everything that influenced its judgement \cite{turpin2023language}. Because the agents searched the open web rather than a fixed set of documents, they also cannot show how representative the sources were of the evidence available, and we did not check whether the financial claims in the reports were correct. Nor did we ask users which criteria they had in mind, so we cannot say whether any report judged the companies the way a given user would want.

\section{Conclusion}
This paper examined how web research agents interpret a stated priority, and which criteria their reports use when the user does not supply them. We studied two agents that compared Microsoft and Alphabet as businesses under a growth priority, a resilience priority or none, with each agent running each condition six times. Most reports took up the requested priority, but the two agents judged it on different grounds. On growth, this led to opposite answers. On resilience, it led to the same answer for partly different reasons. The Anthropic agent's search wording differed significantly by priority. Knowing which priority a report focused on is not enough to know how it compared the companies. Evaluators need to check the criteria a report uses for each judgement. Interfaces could also let users pick those criteria at the start of a task, rather than leaving the choice to the agent.

\section*{Data Availability}
The prompts, run records and collection code are available at \newline \url{https://github.com/Parisasuchdev/user_specified_priorities}.

\section*{Generative AI Usage Statement}
The authors used generative AI tools (Claude and ChatGPT) in the preparation of this manuscript. These tools assisted with polishing author-written text, formatting LaTeX and references, summarizing related work, and checking numbers and quotes against the run records. They were not used to design the study, label the data, run the analysis or determine the findings. The authors checked every number, quote and citation against its source, edited all outputs and remain fully accountable for this work.

\bibliographystyle{ACM-Reference-Format}
\bibliography{references}
\appendix
\section{Agents and Recorded Data}
\label{app:agents}

All 36 runs were submitted through each provider's batch service on 12 October 2026 at 22:25 UTC and finished without an error.

Across its 18 runs, the OpenAI agent issued 247 search queries and the Anthropic agent issued 203. One search by the OpenAI agent could contain several queries, and we counted each one separately, while the Anthropic agent sent one query per search. Their searches returned 1,777 and 1,890 source records respectively, and one URL could appear in several records. The OpenAI reports contained 313 citations and the Anthropic reports 961. The Anthropic agent had web fetch available in every run but never called it.

\section{Statistical Tests}
\label{app:tests}
Tests across all three conditions shuffled the condition labels 5,000 times, and their $p$ values count the observed result as one of the shuffles. Tests between two conditions involved 12 runs, so we computed them exactly over all 924 ways to split the runs into two groups of six. The overlap tests were one-sided, because only greater overlap within a condition points to a difference. The query-share tests were two-sided. Table 4 lists all 46 tests.
\begin{table}
  \caption{All 46 tests, unadjusted $p$ / Holm-adjusted $p$. G is growth, R is resilience and N is neutral.}
  \label{tab:tests}
  \begin{tabular*}{\linewidth}{@{\extracolsep{\fill}}llll@{}}
    \toprule
    Measure & Compared & OpenAI & Anthropic \\
    \midrule
    Report words & All & .0008 / .035 & .0004 / .018 \\
     & G vs R & .0065 / .24 & .0022 / .09 \\
     & G vs N & .089 / 1 & .15 / 1 \\
     & R vs N & .0022 / .09 & .0043 / .16 \\
    \addlinespace
    Query words & All & .52 / 1 & .0002 / .009 \\
     & G vs R & .14 / 1 & .0022 / .09 \\
     & G vs N & .47 / 1 & .0022 / .09 \\
     & R vs N & .85 / 1 & .0022 / .09 \\
    \addlinespace
    Returned URLs & All & .52 / 1 & .0064 / .24 \\
     & G vs R & .20 / 1 & .017 / .59 \\
     & G vs N & .43 / 1 & .020 / .64 \\
     & R vs N & .91 / 1 & .24 / 1 \\
    \addlinespace
    Cited URLs & All & .49 / 1 & .064 / 1 \\
     & G vs R & .33 / 1 & .026 / .81 \\
     & G vs N & .77 / 1 & .17 / 1 \\
     & R vs N & .33 / 1 & .29 / 1 \\
    \addlinespace
    Share of & G vs R & .061 / 1 & .024 / .76 \\
    growth & G vs N & .64 / 1 & .16 / 1 \\
    queries & R vs N & .18 / 1 & .36 / 1 \\
    \addlinespace
    Share of & G vs R & .030 / .91 & .048 / 1 \\
    resilience & G vs N & .087 / 1 & .089 / 1 \\
    queries & R vs N & .83 / 1 & .24 / 1 \\
    \addlinespace
    Cited URLs & All & 1 / 1 & .011 / .38 \\
    unique to one & & & \\
    condition & & & \\
    \bottomrule
  \end{tabular*}
\end{table}

\section{ Coding Rules}
\label{app:coding}
A growth or resilience ranking counted only when the report used the named word to rank the two companies. For growth, that meant growth, grow or growing applied to future company-level prospects. For resilience, it meant resilience, resilient or resiliency. Current growth, a single segment or a claim about upside did not count as a growth ranking, and safety or diversification did not stand in for resilience. A report that gave different winners on different grounds without choosing one was labeled not stated. Unclear was kept for rankings with an ambiguous direction or scope. The safer and upside labels likewise needed safe, predictable or upside in a direct comparison of the two companies. Coders ignored planning text, copied prompts and opinions a report quoted without adopting.

Of the 450 queries, 64 were labeled growth, 60 resilience and 326 neither, and none received both labels. For example, ``Waymo expansion 2026 robotaxi revenue cities'' was labeled growth, ``Alphabet credit rating Moody's S\&P Aa2 AA+ 2025'' was labeled resilience and ``Alphabet 2025 annual report 10-k investor relations'' received neither.

Before the main coding, the two coders labeled a pilot set of 34 search queries and sentences from reports and cited sources with an earlier version of the labels. Agreement was $\kappa = .88$ for growth and $\kappa = .81$ for resilience. In the main sample, the three report disagreements were on overall preference in OpenAI resilience run 02, growth comparison in Anthropic neutral run 05 and report orientation in OpenAI neutral run 03.

\section{Rankings and Criteria by Run}
\label{app:criteria}

Table 5 lists, for every run, the company each report ranked higher and the criteria behind its rankings. A criterion counted only when the report used it to support a judgement about the two companies, so a passage describing one company alone did not count. Both agents also pointed to recurring revenue, switching costs and regulation. We did not classify these, because they did not separate the agents.

\begin{table*}[t]
  \caption{Rankings and criteria for all 36 runs. Overall, Growth and Resilience give the company each report explicitly ranked higher. Cycles means exposure to economic and advertising cycles. A dash means no such ranking or neither examined criterion. Runs with two criteria have one line per criterion.}
  \label{tab:criteria}
  \footnotesize
  \begin{tabular}{@{}p{0.085\textwidth}p{0.07\textwidth}p{0.07\textwidth}p{0.075\textwidth}p{0.09\textwidth}p{0.51\textwidth}@{}}
    \toprule
    Run & Overall & Growth & Resilience & Criterion & Quote \\
    \midrule
    \multicolumn{6}{@{}l}{\textit{OpenAI agent}} \\
    Neutral 01 & -- & -- & -- & -- & \\
    Neutral 02 & Microsoft & -- & -- & -- & \\
    Neutral 03 & Microsoft & -- & Microsoft & Not classified & ``If assessing business resilience and predictability, I would favor Microsoft.'' \\
    Neutral 04 & -- & -- & -- & -- & \\
    Neutral 05 & -- & -- & -- & -- & \\
    Neutral 06 & -- & -- & -- & -- & \\
    Growth 01 & Microsoft & Unclear & -- & -- & \\
    Growth 02 & -- & Microsoft & -- & Risk & ``Microsoft has the stronger risk-adjusted long-term growth profile.'' \\
    Growth 03 & Microsoft & -- & -- & -- & \\
    Growth 04 & -- & Microsoft & -- & Risk & ``Microsoft has the superior risk-adjusted long-term growth profile.'' \\
     & & & & Cycles & ``a macro downturn or a structural decline in search-ad economics would matter more to Alphabet'' \\
    Growth 05 & -- & Microsoft & -- & Predictability & ``If selecting strictly on quality and predictability of long-term business growth, I would place Microsoft modestly ahead.'' \\
    Growth 06 & Microsoft & Microsoft & -- & Risk & ``Microsoft has the stronger risk-adjusted five-year growth profile'' \\
    Resilience 01 & Microsoft & -- & Microsoft & Cycles & ``Advertising normally suffers more in recessions than enterprise subscriptions'' \\
    Resilience 02 & -- & -- & Microsoft & Cycles & ``\ldots tied more closely to enterprise productivity, security, infrastructure, and compliance budgets than to advertising cycles.'' \\
    Resilience 03 & -- & -- & Microsoft & Cycles & ``advertising budgets can weaken in recessions'' \\
    Resilience 04 & -- & -- & Microsoft & Cycles & ``less cyclical and less dependent on consumer attention or advertising budgets'' \\
    Resilience 05 & -- & -- & Microsoft & Cycles & ``advertising budgets are more cyclical than enterprise software commitments'' \\
    Resilience 06 & -- & -- & Microsoft & Cycles & ``Advertising also tends to be more economically sensitive than enterprise subscriptions'' \\
    \midrule
    \multicolumn{6}{@{}l}{\textit{Anthropic agent}} \\
    Neutral 01 & -- & -- & -- & -- & \\
    Neutral 02 & -- & Alphabet & -- & Growth rate & ``Alphabet's top-line growth rate has now overtaken Microsoft's'' \\
     & & & & Cycles & ``making Alphabet more exposed to macro ad cycles'' \\
    Neutral 03 & -- & -- & -- & -- & \\
    Neutral 04 & -- & -- & -- & -- & \\
    Neutral 05 & -- & -- & -- & -- & \\
    Neutral 06 & -- & -- & -- & Cycles & ``heavier reliance on advertising cyclicality'' \\
    Growth 01 & -- & Alphabet & -- & Growth rate & ``Alphabet currently shows the stronger growth inflection'' \\
    Growth 02 & -- & -- & -- & -- & \\
    Growth 03 & -- & -- & -- & -- & \\
    Growth 04 & -- & -- & -- & -- & \\
    Growth 05 & -- & -- & -- & -- & \\
    Growth 06 & -- & Alphabet & -- & Growth rate & ``Alphabet's growth rate may remain structurally higher over the next five years'' \\
    Resilience 01 & -- & -- & -- & Cycles & ``more cyclically exposed to ad-spend downturns'' \\
     & & & & Credit ratings & ``without touching its AAA credit rating'' \\
    Resilience 02 & -- & -- & Microsoft & Credit ratings & ``Microsoft holds AAA credit ratings from S\&P Global, Moody's, and Fitch'' \\
    Resilience 03 & -- & -- & -- & -- & \\
    Resilience 04 & -- & -- & -- & -- & \\
    Resilience 05 & -- & -- & Microsoft & Credit ratings & ``Microsoft holds AAA credit ratings from S\&P, Moody's, and Fitch'' \\
    Resilience 06 & -- & Alphabet & Microsoft & Growth rate & ``Alphabet screens as the higher-growth, somewhat higher-risk position'' \\
     & & & & Credit ratings & ``carries the highest possible credit rating'' \\
    \bottomrule
  \end{tabular}
\end{table*}
\end{document}